\documentclass{elsart}
\usepackage{latexsym,amssymb,amsmath,amsfonts}
\usepackage{epsfig}

\makeatletter \@addtoreset{equation}{section} \makeatother
\renewcommand{\theequation}{\thesection.\arabic{equation}}
\newcommand {\cp}        {\cos\varphi}

\renewcommand {\sp}      {\sin\varphi}

\newcommand {\Exp}    [1]{\exp\left\{{#1}\right\}}
\newcommand {\vvec}   [2]{\begin{pmatrix}{#1}\\{#2}\end{pmatrix}}

\newcommand {\THETA}[2]  {\vartheta\begin{bmatrix}{#1}\\{#2}\end{bmatrix}}
\newcommand {\rmd}       {{\rm d}}

\newcommand {\rhs}       {right-hand side}
\newcommand {\bcs}       {boundary conditions}
\newcommand {\CS}        {Chern--Simons}

\newcommand {\gsim}{\mathrel{\hbox{\rlap{\lower.55ex \hbox {$\sim$}}
            \kern-.3em \raise.4ex \hbox{$>$}}}}
\newcommand {\lsim}{\mathrel{\hbox{\rlap{\lower.55ex \hbox {$\sim$}}
            \kern-.3em \raise.4ex \hbox{$<$}}}}

\newcommand {\refeq}[1] {(\ref{#1})}
\newcommand {\beq}[0]{\begin{equation}}
\newcommand {\eeq}[0]{\end{equation}}

\renewcommand {\bar}     {\overline}
\def\id{{\rm 1\kern-.12em
        \rule{0.3pt}{1.5ex}\raisebox{0.0ex}{\rule{0.1em}{0.3pt}}}}

\begin{document}
\noindent  hep-th/0105310                              \hfill  KA--TP--17--2001
\vspace*{1cm}
\runauthor{Klinkhamer and Mayer}
\begin{frontmatter}

\title{Torsion, topology and CPT anomaly in two-dimensional chiral $U(1)$
       gauge theory}

\author{F.R.~Klinkhamer},
\ead{frans.klinkhamer@physik.uni-karlsruhe.de}
\author{C.~Mayer}
\ead{cm@particle.uni-karlsruhe.de}

\address{Institut f\"ur Theoretische Physik, Universit\"at
Karlsruhe, D--76128 Karlsruhe, Germany}

\begin{abstract}
  We consider the CPT anomaly of two-dimensional
  chiral $U(1)$ gauge theory on a torus with topologically nontrivial
  zweibeins corresponding to the presence of spacetime torsion.
  The resulting chiral determinant can be expressed in terms of the
  standard chiral determinant without torsion, but with modified spinor boundary
  conditions. This implies that the two-dimensional CPT anomaly can be moved from one
  spin structure to another by choosing appropriate zweibeins.
  Similar results apply to higher-dimensional chiral gauge theories.
\end{abstract}

\end{frontmatter}

\section{Introduction}

Recently, it has been shown that there is a violation of CPT invariance in
certain (non-)Abelian chiral gauge theories defined on nonsimply connected spacetime
manifolds \cite{K00}.
The well-known CPT theorem \cite{P55,L57} is evaded by the breakdown of
Lorentz invariance due to the quantum effects of the chiral fermions \cite{K00,K98}.

The Abelian CPT anomaly is particularly obvious on the two-di\-men\-sion\-al
torus \cite{KN01} where the chiral determinant can be
computed exactly (see Ref.~\cite{IN99} and references therein).
For this reason, we will consider in this paper primarily two-di\-men\-sion\-al
chiral $U(1)$ gauge theory defined over the torus.
More precisely, we study the CPT anomaly on the torus in the presence of spacetime
torsion. That is, we consider the effects of
a nontrivial con\-fig\-u\-ra\-tion of zweibeins,
which gives rise to a nonvanishing torsion tensor \cite{HHKN76,EGH80}.
(Zweibeins are the two-dimensional analogs of vierbeins or tetrads
in four dimensions.)

The main goal of the present paper is then to understand better the role of topology
and spin structure for the two-dimensional CPT anomaly, by studying the
response of the chiral determinant to the introduction of topologically nontrivial
zweibeins on the torus. For a general discussion of spinors over nonsimply connected
spacetime manifolds, we refer the reader, in particular, to Refs.~\cite{AI79,BD79}.

The paper is organized as follows. In Section~2, we discuss some aspects of
the geometry of the two-dimensional torus with torsion and establish our notation.
Specifically, we mention two consequences of torsion at the level
of the spacetime structure. Namely,
parallelograms need not close and extremal and autoparallel curves need not coincide.
We also comment on some interesting properties of  topologically nontrivial
zwei- (or vier-)beins on the torus and their possible origin.

In Section~3, we show that one can relate a fermionic
Lagrangian with nontrivial zweibeins to a
Lagrangian with trivial zweibeins by a simple spinor redefinition.
This field redefinition can, however, change the spinor boundary conditions.

In Section~4, we use this property of the fermionic
Lagrangian to express the chiral determinant for topologically
nontrivial zweibeins in terms of the chiral determinant for
trivial zweibeins but modified  spinor boundary conditions.
We also give a heuristic argument for the result.
A similar calculation is done in Appendix A for the Dirac determinant of a
vector-like $U(1)$ gauge theory.

In Section~5, we discuss the role of torsion for the two-dimensional chiral CPT
anomaly and find that only the topologically nontrivial part of the zweibeins
affects the anomaly.

In Section~6, we summarize our results and briefly comment on the four-dimensional case.

\section{Topology, geometry and torsion}

\subsection{Zweibeins on the torus}

The Cartesian coordinates $x^\mu\in [0,L]\;,\; \mu=1,2,\;$ are taken to parameterize a
particular two-dimensional torus
$T^{2}[i]$, with modulus (Teichm\"uller parameter) $\tau=i$.
This torus can be thought of as a square with
flat Euclidean metric $g_{\mu\nu}(x)=\delta_{\mu\nu}\equiv\text{diag}(1,1)$ and
opposite sides identified; see Fig.~1.

Zweibeins locally define an orthonormal basis of one-forms
\begin{equation}\label{orthonormal}
  e^a(x) = e^a_\mu(x)\,\rmd x^\mu\;.
\end{equation}
Here, Latin indices ($a$, $b$, $\ldots$) refer to the local frame and Greek indices
($\mu$, $\nu$, $\ldots$) to the base space.
Throughout this paper, summation over equal upper and lower indices is understood.
The inverse zweibeins $ e^\mu_a(x)$ are defined by
\begin{equation}\label{delta=ee}
  \delta_a^{\,\;b} =  e^\mu_a(x)\;e_\mu^b(x)\;.
\end{equation}
In terms of the zweibeins, the metric can be expressed as follows
(see, for example, Refs.~\cite{HHKN76,EGH80}):
\begin{equation}\label{g=ee}
  g_{\mu\nu}(x) = e_\mu^a(x)\;e_\nu^b(x)\;\delta_{ab}\;.
\end{equation}
Inversely, this equation defines the zweibeins, but only up to a space-dependent
orthogonal transformation.
\begin{figure*}[t]\label{fab}
 \centerline{\epsfysize=1.75in \epsfbox{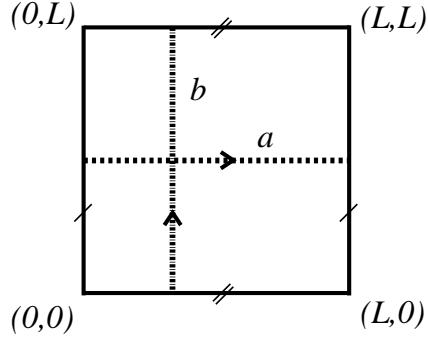}}
  \caption{The torus $T^{2} [i]$
  is represented as a square with opposite sides identified.
  Two distinct noncontractible curves are labeled $a$ and $b$.}
\vspace*{1cm}
\end{figure*}

In the following, we consider zweibeins  $e^a_\mu(x)$ taking
values in a matrix representation of the group $SO(2)$:
\begin{equation}\label{zb}
  \vvec{e^1_\mu(x)}{e^2_\mu(x)} \;\equiv\; \delta_\mu^1
  \begin{pmatrix}\cp(x)\\\sp(x)\end{pmatrix}+\delta_\mu^2
  \begin{pmatrix}-\sp(x)\\\ \ \;\cp(x)\end{pmatrix}\;,
\end{equation}
parametrized by the real function $\varphi(x)$. The corresponding
metric is flat, $g_{\mu\nu}(x)$ $=$ $\delta_{\mu\nu}$.
Obviously, the choice $\varphi(x)=0$ yields the trivial zweibeins
$e_\mu^a(x)=\delta_\mu^a$.
The extra degree of freedom $\varphi(x)$ in the zweibeins (\ref{zb})
can, however, generate torsion, as will be shown in the next subsection.

The zweibeins \refeq{zb} need to be defined in a consistent way on $T^{2} [i]$,
namely
\begin{equation}\label{zweibeinbcs}
  e_\mu^a(0,x^2) = e_\mu^a(L,x^2)\;,\qquad e_\mu^a(x^1,0) =
  e_\mu^a(x^1,L)\;.
\end{equation}
This requirement naturally leads to a decomposition of the $SO(2)$ rotation
angle $\varphi(x)$ into topologically trivial and nontrivial parts:
\begin{equation}\label{varphi}
  \varphi(x) \equiv  \omega(x) + \chi(x) \;,
\end{equation}
where the real function $\omega(x)$ is taken to be strictly periodic in $x^1$ and $x^2$,
with period $L$.
The other real function $\chi(x)$ is associated with the two generating
curves $a$ and $b$ of the homology group $\mathcal H^1(T^{2},\mathbb Z)$
$=$ $\mathbb Z \oplus \mathbb Z$; cf. Ref.~\cite{EGH80}.
Specifically, the function $\chi(x)$ is
given by
\begin{equation}\label{mn}
  \chi(x) \equiv  (2\pi/L)(m\,x^1+n\,x^2)\;,\qquad
  m,n\in\Zset\;,
\end{equation}
for the two distinct noncontractible curves $a$ and $b$ shown in Fig.~1.
[The notation $\chi(x;m,n)$ would, of course, be more accurate.]

\subsection{Connection}

The condition for parallel transport of an arbitrary vector field $C^\mu(x)$ along the
infinitesimal path $(x,x+\delta x)$ is
\begin{equation} \label{parallelC}
  C^\lambda(x)\, e_\lambda^a(x) =
  \big(C^\lambda(x) + \delta C^\lambda(x) \big)\,e_\lambda^a(x+\delta x)\;.
\end{equation}
With the linear \emph{Ansatz} \cite{E55}
\begin{equation}\label{conn_def}
  \delta C^\lambda(x) \;\equiv\; -\Hat\Gamma^\lambda_{\mu\nu}(x)\,
  C^\mu(x)\,\delta x^\nu\;,
\end{equation}
condition \refeq{parallelC} enables us to express the Riemann--Cartan connection
$\Hat\Gamma^\lambda_{\mu\nu}(x)$ in terms of the zweibeins:
\begin{equation}\label{conn}
  \Hat\Gamma^\lambda_{\mu\nu}(x) = e^\lambda_a(x)\,e^a_{\mu,\nu}(x) =
  (\delta^\lambda_2\delta_\mu^1 - \delta^\lambda_1\delta_\mu^2)\:\partial_\nu\varphi(x)\;,
\end{equation}
for the particular zweibeins \refeq{zb} which have vanishing metric
Christoffel symbol \cite{EGH80,AI79}.
As usual, the notation $\phi_{,\nu}$ stands for $\partial\phi/\partial x^\nu$.
The result \refeq{conn} demonstrates that the freedom in choosing a particular connection
on $T^{2}[i]$ is a consequence of the fact that for a fixed metric
$g_{\mu\nu}(x)$ $=$ $\delta_{\mu\nu}$ the zweibeins are defined only up to an
orthogonal transformation.

Since the zweibeins considered take values in a two-dimensional matrix
representation of the group $SO(2)$, we can write the transformation of the connection
under an $SO(2)$ transformation $\Tilde e_a^\mu(x)$ as
\begin{equation}\label{conn_trf}
  \Hat\Gamma^{\,\prime \, a}_{\,\ \mu b} = \Tilde e_\lambda^a\,\Hat\Gamma^\lambda_{\mu\nu}\,\Tilde e^\nu_b +
  (\partial_\mu\Tilde e^a_\nu)\, \Tilde e_b^\nu\;.
\end{equation}
While the connection itself does not transform as a tensor, its antisymmetric component in
the two lower indices does. This object is called the torsion tensor \cite{HHKN76,EGH80},
\begin{equation}
  T^\lambda_{\mu\nu}(x) \equiv \Hat\Gamma^\lambda_{\mu\nu}(x)-\Hat\Gamma^\lambda_{\nu\mu}(x)\;.
\end{equation}
For the connection \refeq{conn}, we obtain
\begin{equation}\label{torsion}
  T^\lambda_{12}(x) = -T^\lambda_{21}(x) = \partial^\lambda\varphi(x)\;,
\end{equation}
where the function $\varphi(x)$ parameterizes the zweibeins (\ref{zb}).

Parallel transport enables us to give an ``operational definition'' of
torsion (see also Refs.~\cite{HHKN76,E55}).
Consider the parallelogram spanned by the line
elements $(x,E_1\cdot x)$ and $(x,E_2\cdot x)$,
where the infinitesimal displacements  of $x$ are given by
\begin{equation}
  \begin{split}
    (E_1\cdot x)^\mu &= x^\mu + \epsilon_1^\mu(x)\;, \quad
    \epsilon_1^\mu(x) \;\equiv\; \epsilon_1^a\, e_a^\mu(x)\;,\\[0.25\baselineskip]
    (E_2\cdot x)^\mu &= x^\mu + \epsilon_2^\mu(x)\;,\quad
    \epsilon_2^\mu(x) \;\equiv\; \epsilon_2^a\, e_a^\mu(x)\;,
  \end{split}
\end{equation}
with real infinitesimal coefficients $\epsilon_1^a$ and $\epsilon_2^a$.
There is then torsion if the
parallelogram does not close, i.e.~$[E_1E_2$ $-$ $E_2E_1]$ $\cdot$ $x\neq 0$.
In fact, a short calculation yields
\begin{eqnarray}\label{deficit}
  \left(\,[E_2 E_1-E_1 E_2]\cdot x\right)^\mu &=&
  [\,\Hat\Gamma^\mu_{\rho\sigma}(x) - \Hat\Gamma^\mu_{\sigma\rho}(x)\,]\,
  \epsilon_2^\rho(x)\,\epsilon_1^\sigma(x)\nonumber\\[0.25\baselineskip]
  &=& T^\mu_{\rho\sigma}(x)\;\epsilon_2^\rho(x)\,\epsilon_1^\sigma(x)\;,
\end{eqnarray}
where Eq.~\refeq{conn_def} has been used for the parallel transport of
$\epsilon_1^\mu(x)$ and $\epsilon_2^\mu(x)$.

\subsection{Extremal and autoparallel curves}

Apart from the fact that parallelograms do not close in the presence of torsion, there
is a further consequence of torsion at the level of the spacetime structure: extremal
and autoparallel curves do not necessarily coincide (see also Ref.~\cite{HHKN76}).

The equations for an extremal curve (shortest or longest line)
on the two-dimensional  flat spacetime manifold $T^{2} [i]$ are given by
\begin{equation}\label{ex}
  \ddot x^1(\tau) \;=\; 0\;,\qquad \ddot x^2(\tau) \;=\; 0\;,
\end{equation}
where the dot denotes differentiation with respect to the affine parameter $\tau$.
Extremal curves on $T^{2} [i]$ may or may not close.
An example of an extremal curve that does \emph{not} close is given by
$x^1=\tau$ and $x^2=\sqrt{2}\,\tau$, since $\sqrt{2}$ is an irrational number.
Recall that the ratio of the periodicities in $x^1$ and $x^2$ is exactly $1$
for the particular torus $T^{2} [i]$ considered; see Fig.~1.

The equations for an autoparallel curve (straightest line)
can be deduced by requiring that autoparallel
curves are always tangent to the zweibeins.  The substitution $C^\mu=\rmd
x^\mu(\tau)/\rmd \tau \equiv \dot x^\mu(\tau)$ in Eq.~\refeq{conn_def} gives the following
result:
\begin{equation}
  \ddot x^\lambda + \Hat\Gamma^\lambda_{\mu\nu}(x)\,\dot x^\mu\dot x^\nu \;=\; 0\;.
\end{equation}
Note that the Riemann--Cartan connection $\Hat\Gamma^\lambda_{\mu\nu}$ in
Eq. (\ref{conn}) also has a symmetric part in $\mu\nu$.
For the special case of $\omega(x)=0$ in Eq.~\refeq{varphi}, the equations become
\begin{equation}\label{auto}
  \begin{split}
    \ddot x^1 - (2\pi/L)\,(m\,\dot x^1 + n\,\dot x^2)\,\dot x^2 &\;=\;0\;,\\[0.25\baselineskip]
    \ddot x^2 + (2\pi/L)\,(m\,\dot x^1 + n\,\dot x^2)\,\dot x^1 &\;=\;0\;.
  \end{split}
\end{equation}
Since the general solution of these coupled differential equations can be quite involved,
we only discuss the class of solutions satisfying
\begin{equation}\label{rf}
  m\,\dot x^1 + n\,\dot x^2 \;=\; 0\;.
\end{equation}
In this case, the Eqs.~\refeq{auto} reduce to Eqs.~\refeq{ex}, but with the additional
constraint \refeq{rf} on the constant velocities $\dot x^\mu$.
One easily recognizes that Eq. (2.19) describes autoparallel curves
by noting that solutions to this equation are curves of constant $\chi$; see
Eq. (\ref{mn}).

In contrast to extremal curves, which may or may not close on the torus $T^{2} [i]$,
the autoparallel curves given by Eqs.~\refeq{auto}-\refeq{rf}
always close, since their slopes are rational
($\dot x^2/\dot x^1=-m/n$).
More specifically, the extremal curve $(x^1, x^2)$ $=$ $(\tau,\sqrt{2}\,\tau)$
mentioned a few lines below Eq. (\ref{ex}) does \emph{not} solve the Eqs. (\ref{auto})
and is, therefore, not autoparallel, as long as the torsion parameter
function is topologically nontrivial, $\varphi(x)$  $=$ $\chi(x)$ $\neq$ $0$.
This clearly shows the difference of the two types
of curves in the presence of torsion.

\subsection{Topologically nontrivial vielbeins}

In this last subsection on geometry, we elaborate on the special nature of
topologically nontrivial zweibeins
\refeq{zb}--\refeq{mn} with $(m,n)$ $\neq$ $(0,0)$ and $\omega(x)$ $=$ $0$.
For these zweibeins, namely, the torsion tensor \refeq{torsion}
would be constant (and nonzero) over the whole spacetime manifold $T^2[i]$. This
would then correspond to a new local property of spacetime. One manifestation
would be the nonclosure of parallelograms as discussed in Section 2.2.

For topologically nontrivial zweibeins \refeq{zb}--\refeq{mn} with $(m,n)$ $=$ $(1,1)$
and $\omega(x)$ $=$ $0$, a typical
parallelogram obtained by parallel transport,
with lengths $|\epsilon_1^\mu|$ $=$ $|\epsilon_2^\mu|$ $=$ $\ell$,
would fail to close by a distance \refeq{deficit} of order
\begin{equation}\label{orderdeficit}
 (2\pi /L) \: \ell^{\,2} \sim 10^{-25}\, \mathrm{m}\;
 ( 10^{10}\,\mathrm{lyr}/L)\, ( \ell/\mathrm{m})^2 \;,
\end{equation}
which would still be 10 orders of magnitude above the Planck length
$l_{\mathrm{P}}$ $\equiv$ $\sqrt{\hbar\,G/c^3}$.
As will be discussed in Section 6, a similar torsion effect may occur in four
(or more) dimensions, for appropriate vier- (or viel-)beins.
The level of accuracy indicated by Eq. \refeq{orderdeficit}
might be, in principle, within reach of experiment. [We have in mind a rapidly
rotating ($\sim 100$ Hz) experimental setup in a free-fall environment (e.g. in a
drag-free satellite). See, for example, Eq. (3.35) in Ref. \cite{DB79} for
the optimal sensitivity of a resonant-bar detector for periodic gravitational
waves.]

Throughout this paper,  we consider the zwei- or vierbeins as fixed
classical background fields. Let us, however, briefly
remark on the possible origin of translation-invariant
torsion resulting from topologically nontrivial vierbeins (see also Section~6).
The crucial point is that this type of torsion would not
have to be generated dynamically by a local spin density, but
could perhaps arise as a kind of boundary condition (most likely, set at the beginning
of our universe).\footnote{\label{ftn1}Note that the
two-dimensional gravitational Einstein--Cartan action
\cite{HHKN76}, based on the Ricci scalar defined in terms of the connection
\refeq{conn}, gives the field equation $\partial^2 \varphi$ $=$ $0$, which is trivially
solved by the configuration \refeq{mn}. The situation in four dimensions is less
satisfactory, as will be discussed in Section 6.}
Moreover, the spin density can only be expected to
give a negligible contribution to the torsion
tensor for the present cosmological number densities $n$ of protons or electrons.
In fact, the order of magnitude to be compared with Eq.
\refeq{orderdeficit} is
\begin{equation}
(G\,c^{-3})\,(\hbar\,n)\, \ell^2 \sim
n\,l_{\mathrm{P}}^2 \,\ell^2 \sim
10^{-70}\,\mathrm{m} \;(n/\mathrm{m}^{-3}) \,(\ell/\mathrm{m})^2\;.
\end{equation}
(See also Section V A 3 of Ref. \cite{HHKN76}.)
Hence, the translation-invariant torsion from topologically nontrivial vierbeins
may at present be an extremely weak effect, but the effect is still many orders of
magnitude above that expected from the ordinary matter of the universe.

\section{Fermionic Lagrangian}

We use a ``chiral'' basis for the two-dimensional Dirac matrices
\begin{equation}\label{gamma}
  \gamma^1 \;\equiv\; \begin{pmatrix}0&+1\\+1&0\end{pmatrix}\,,\quad
  \gamma^2 \;\equiv\;  \begin{pmatrix}0&+i\\-i&0\end{pmatrix}\,,\quad
  \gamma_\mathrm{S} \;\equiv\; i \gamma^1\gamma^2 =
  \begin{pmatrix}+1&0\\0&-1\end{pmatrix}\,,
\end{equation}
where $\gamma_\mathrm{S}$ anticommutes with $\gamma^1$ and $\gamma^2$.
(The suffix S stands for ``strong reflection,'' originally introduced by Pauli
for the proof of the CPT theorem \cite{P55,L57}.)

For trivial zweibeins $e^a_\mu(x)=\delta^a_\mu\equiv\id$, the manifestly Hermitian
Lagrangian is given by
\begin{equation}\label{l}
  \mathcal L[\Bar\Psi,\Psi,A,\id] = (i/2)\,\Bar\Psi\,
  \gamma^\mu(\overset{\rightarrow}{\partial_\mu} +iA_\mu)\,\Psi -
  (i/2)\,\Bar\Psi\,(
  \overset{\leftarrow}{\partial_\mu} - iA_\mu)\gamma^\mu\, \Psi\;,
\end{equation}
with constant gamma matrices
\begin{equation}
\gamma^\mu \,\equiv\, \delta_a^\mu\,\gamma^a.
\end{equation}
This Lagrangian is invariant
under global $SO(2)$ transformations (with $\gamma_\mathrm{S}$ as defined above),
\begin{equation} \label{rigidSO(2)}
  \Psi(x) \;\longrightarrow\; e^{-i\kappa\gamma_\mathrm{S}/2}\,\Psi(x')\;,\quad
  \Bar\Psi(x)\;\longrightarrow\; \Bar\Psi(x')\,e^{i\kappa\gamma_\mathrm{S}/2},
\end{equation}
and local $U(1)$ gauge transformations,
\begin{equation} \label{gaugeU(1)}
  \Psi(x) \;\longrightarrow\; \e^{i\xi(x)}\,\Psi(x)\;,\quad
  \Bar\Psi(x)\;\longrightarrow\; \Bar\Psi(x)\,\e^{-i\xi(x)},
\end{equation}
supplemented by the usual transformations of the gauge field $A_\mu(x)$.
For the basis of gamma matrices \refeq{gamma}, the two (independent) Dirac
spinors can be decomposed into four one-component Weyl spinors:
\begin{equation}
  \Psi(x)\equiv \begin{pmatrix}\psi_R(x)\\\psi_L(x)  \end{pmatrix}\;,\quad
  \Bar\Psi(x)\equiv \begin{pmatrix}\Bar\psi_R(x) &\Bar\psi_L(x)  \end{pmatrix}\;,
\end{equation}
where $(1\mp\gamma_\mathrm{S})/2$ projects on the left- and right-moving subspaces
of solution space.

Throughout this paper, we consider only topologically trivial gauge potentials $A_\mu(x)$.
We therefore take the $U(1)$ gauge potential $A_\mu(x)$ to be periodic in $x^1$ and $x^2$,
with period $L$. The spinors are allowed to have either periodic or antiperiodic boundary
conditions:
\begin{equation}\label{bcs}
  \begin{split}
    \Psi(x^1+L,x^2) &= -\e^{2\pi i\theta_1}\;\Psi(x^1,x^2)\;,\\[0.25\baselineskip]
    \Psi(x^1,x^2+L) &= -\e^{2\pi i\theta_2}\;\Psi(x^1,x^2)\;.
  \end{split}
\end{equation}
(The adjoint spinors $\Bar\Psi(x)$ obey the same \bcs.)
The variables $\theta_1,\theta_2 \in \{0,1/2\}$ then fix the
spinor boundary conditions, with $(\theta_1,\theta_2)$  $=$
$(1/2,1/2)$ corresponding to doubly-periodic
boundary conditions and $(\theta_1,\theta_2)$  $=$
$(0,0)$ to doubly-antiperiodic boundary conditions.
Mixed spinor boundary conditions correspond to $(\theta_1,\theta_2)$ $=$
$(1/2,0)$ or $(0,1/2)$. The four possible combinations of
$(\theta_1,\theta_2)$ are said to define the four spin structures over the torus.

For the general zweibeins \refeq{zb}, the Lagrangian \refeq{l} becomes
\begin{equation}\label{le}
  \mathcal L[\Bar\Psi,\Psi,A,e] =
  (i/2)\,\Bar\Psi\,\Hat\gamma^\mu D_\mu \Psi
  + \mathrm{H.c.}\;,
\end{equation}
with space-dependent gamma matrices
\begin{equation}
\Hat\gamma^\mu(x) \;\equiv\; e_a^\mu(x)\,\gamma^a\;
\end{equation}
and covariant derivatives
\begin{equation}
 D_\mu\Psi \;\equiv\; (\partial_\mu+iA_\mu+i\Omega_\mu)\Psi \;.
\end{equation}
The Lagrangian \refeq{le} is invariant under gauged $SO(2)$
transformations  (\ref{rigidSO(2)}) due to the presence of the spin
connection \cite{EGH80,AI79}
\begin{eqnarray}
  \Omega_\mu(x) &\equiv& \Hat\Gamma_\mu^{ab}(x)\,\sigma_{ab}/2 \;=\;
  e_{\nu,\mu}^a e^{b\nu} \,\sigma_{ab}/2 \;=\;
  -\gamma_\mathrm{S}\,\partial_\mu\varphi(x)/2\;,\nonumber\\[0.25\baselineskip]
  \sigma_{ab} &\equiv& i\,[\gamma_a,\gamma_b]/4\;,
\end{eqnarray}
provided Eq.~\refeq{conn_trf} is used for the transformations.
Here, the generator of $SO(2)$ transformations is
given by $\sigma_{12} = -\sigma_{21} = \gamma_\mathrm{S}/2$, for the
representation of gamma matrices chosen.

It is now possible to rewrite the Lagrangian $\refeq{le}$ as follows:
\begin{equation}\label{l_le}
    \mathcal L[\Bar\Psi,\Psi,A,e] =
    \mathcal L\left[\Bar\Psi \e^{i\varphi\gamma_\mathrm{S}/2},
    \e^{-i\varphi\gamma_\mathrm{S}/2}\Psi,A,\id\right]\;,
\end{equation}
for zweibeins $e^a_\mu(x)$ and parameter function
$\varphi(x)$ given by Eqs.~(\ref{zb}) and \refeq{varphi},
respectively. The effects of the nontrivial zweibeins (\ref{zb})
can therefore be absorbed  by the simple spinor redefinition
\begin{equation}\label{redef0}
  \Psi'(x)\;\equiv\; \e^{-i\varphi(x)\gamma_\mathrm{S}/2}\,\Psi(x)\;,\quad
  \Bar\Psi'(x)\;\equiv\; \Bar\Psi(x)\, \e^{i\varphi(x)\gamma_\mathrm{S}/2}\;.
\end{equation}
Obviously,
this field redefinition changes the spinor boundary conditions according to the values
of $m$ and $n$ in the function $\varphi(x)$; see Eqs. (\ref{varphi}) and (\ref{mn}).
Note, however, that this field redefinition is not a proper $SO(2)$ spinor
redefinition on $T^{2} [i]$, in the sense that contractible loops of spinor rotations
need not correspond to contractible loops of coordinate rotations (see Ref.~\cite{AI79}
for further details). Physical consequences of changed spinor boundary conditions are,
for example, the difference of the vacuum energy density \cite{BD79} and
the occurrence of the CPT anomaly (see Section~5).

\section{Chiral determinant with torsion}

In this section, we express the two-dimensional chiral determinant with torsion in terms of
the standard chiral determinant without torsion. This can be done by use of the identity
\refeq{l_le} and field redefinition~\refeq{redef0}.

The chiral determinant with torsion is then given by the following path integral:
\begin{equation}\label{De}
    D^{\{\theta_1,\theta_2\}}[A,e]
    = \int \left[\mathcal D\Bar\psi_R\,\mathcal D\psi_L\right]_{(\theta_1,\theta_2)}\,
    \Exp{-S\left[\Bar\psi_R\, \e^{i\varphi/2},\e^{i\varphi/2}\,\psi_L,A\right]}\;,
\end{equation}
in terms of the standard action for the one-component Weyl spinor $\psi_L(x)$ and its
conjugate $\Bar\psi_R(x)$,
\begin{equation}
  S[\Bar\psi_R,\psi_L,A] = \int_{T^{2} [i]}\rmd^2x\;
  \Bar\psi_R(x)\,\sigma^\mu\big(\partial_\mu+iA_\mu(x)\big)\,\psi_L(x)\;,
\end{equation}
with $\sigma^\mu\equiv(1,i)$.
The parameters $\theta_1$ and $\theta_2$ in Eq.~\refeq{De}
denote the spin\-or boundary conditions for the
compact dimensions; see Eq.~\refeq{bcs}. Recall also that the zweibeins
$e^a_\mu(x)$ are given by Eqs.~(\ref{zb})--(\ref{mn}) and
the corresponding torsion tensor by Eq.~(\ref{torsion}).

The easiest way to calculate the chiral determinant with torsion is to perform the field
redefinition~\refeq{redef0}:
\begin{equation}\label{De0}
    D^{\{\theta_1,\theta_2\}}[A,e]
    = \int \left[\mathcal D(\Bar\psi'_R\,\e^{-i\varphi/2}) \,
                 \mathcal D(\e^{-i\varphi/2}\,\psi'_L)
            \right]_{(\theta'_1,\theta'_2)} \,
            \Exp{-S[\Bar\psi'_R,\psi'_L,A]}\;,
\end{equation}
where $\theta'_1$ and $\theta'_2$ indicate the \bcs~of the transformed spinor fields
(see below). Now, we only need to compute the relevant Jacobians
and the next subsection reviews  a convenient method.

\subsection{Jacobians for infinitesimal phase transformations}

We propose to use Fujikawa's method \cite{F80} to compute the Jacobians of the spinor
redefinition
\refeq{redef0} for the case of an infinitesimal phase $\varphi(x)$ $=$ $\alpha(x)$.
Note, however, that our spinor redefinition is \emph{not} a chiral transformation, as was
the case in Fujikawa's original calculation.

The relevant
Hermitian Dirac operator $(i\FMSlash D) = (i\FMSlash D)^\dagger$ is given by
\begin{equation}
    i\FMSlash{D} \;\equiv\; i\hat\gamma^\mu(x) D_\mu
    \;=\; i\e^{i\varphi\gamma_\mathrm{S}/2}\, \gamma^a \delta^\mu_a
    (\partial_\mu+iA_\mu)\,\e^{-i\varphi\gamma_\mathrm{S}/2}
    \;\equiv\;
    \begin{pmatrix}
      0 & i\FMslash{d}_R \\ i\FMslash{d}_L & 0
    \end{pmatrix}\;,
\end{equation}
with
\begin{equation}
  i\FMslash{d}_R \;\equiv\;  (i\FMslash{d}_L)^\dagger  \;\equiv\;
  \e^{i\varphi/2}\,i\sigma^\mu(\partial_\mu+i A_\mu)\,\e^{i\varphi/2}  \;.
\end{equation}

Since we intend to compute the Jacobians for the left- and right-moving chiral fermions
separately, we can work with the following Hamiltonian:
\begin{equation}\label{H}
  H \equiv (i\FMSlash{D})(i\FMSlash{D})^\dagger = \begin{pmatrix} -\FMslash{d}_R\FMslash{d}_L & 0\\ 0 & -\FMslash{d}_L\FMslash{d}_R
\end{pmatrix} \equiv \begin{pmatrix} H_+&0\\ 0&H_- \end{pmatrix}\;,
\end{equation}
which has the advantage of being Hermitian for each chirality separately, $H_\pm =
(H_\pm)^\dagger$. Explicitly, its components are given by
\begin{equation}\label{H_pm}
  H_\pm = -D_\pm^\mu D_{\pm\,\mu} \,\mp\, F\;,
\end{equation}
with  the further definitions
\begin{equation}\label{Fdef}
D_{\pm\mu}\;\equiv\; \partial_\mu+iA_\mu \mp i\partial_\mu\varphi/2 \;, \quad
F\;\equiv\; \partial_1 A_2-\partial_2 A_1 \;.
\end{equation}

Following Ref.~\cite{F80}, we introduce normalized eigenfunctions of $H_\pm$,
\begin{equation}\label{orthonormality-relations}
  H_\pm\, \phi_{\pm,k}(x) \;=\; \lambda_k^2\;\phi_{\pm,k}(x)\;,\quad
  \int_{T^{2} [i]}\rmd^2x\;\phi_{\pm,k}^\dagger(x)\,\phi_{\pm,l}(x)
  \;=\; \delta_{kl} \;,
\end{equation}
for $k,l\in\Zset$.  The four independent Weyl spinors are then expanded as
follows:
\begin{equation}\label{PsiExpansion}
  \begin{split}
    \psi_R(x) &= \sum_k a_k\,\phi_{+,k}(x)\;,\qquad
    \Bar\psi_R(x) = \sum_k \bar a_k\,\phi^\dagger_{+,k}\;,\\[0.25\baselineskip]
    \psi_L(x) &= \sum_k b_k\,\phi_{-,k}(x)\;,\qquad
    \Bar\psi_L(x) = \sum_k \bar b_k\,\phi^\dagger_{-,k}\;,
  \end{split}
\end{equation}
with Grassmann numbers $a_k,\Bar a_k, b_k, \Bar b_k$. Note that the eigenfunctions
$\phi_{\pm,k}(x)$ have been assigned to the Weyl spinors in order to diagonalize $H$:
\begin{equation}
  <\Bar\Psi |H| \Psi> \;=\; \sum_k \, \lambda_k^2\,(\Bar a_k\, a_k + \Bar b_k\, b_k)\;.
\end{equation}

For the field redefinition used in Eq.~\refeq{De0} and with the definition
\begin{equation}\label{psiRprime}
      \psi'_R(x) \equiv \sum_k a'_k\phi_{+,k}(x)\; ,
\end{equation}
the Grassmann variable $a_k$ changes to
\begin{equation}\label{anprime}
      a'_k = \sum_l \Bigg(\int_{T^{2} [i]}\rmd^2x\;
      \phi^\dagger_{+,k}(x)\,\e^{-i\varphi/2}\,\phi_{+,l}(x)\Bigg)\, a_l
      \equiv \sum_l\,  V_{+,kl}[-\varphi/2]\;a_l\;.
\end{equation}
The changes for the Grassmann variables
$\Bar a_k, b_k$ and $\Bar b_k$ are analogous, but with matrices
$V^{\rm T}_+[\varphi/2]$, $V_-[\varphi/2]$ and $V^{\rm T}_-[-\varphi/2]$
replacing $V_{+}[-\varphi/2]$ in Eq.~(\ref{anprime}). The superscript
$\mathrm{T}$ indicates the transpose of the matrix.

Formally, the functional measure can be written as a product over the
differentials $\rmd\Bar a_k$ and $\rmd b_l$:
\begin{equation}\label{measure}
  \mathcal D\Bar\psi_R\,\mathcal D\psi_L \equiv
  \prod_{k\in\Zset} \rmd\Bar a_k \prod_{l\in\Zset} \rmd b_l\;.
\end{equation}
Under a spinor redefinition, the change of the functional measure is then
given by the corresponding Jacobians,
\begin{equation}\label{pidt}
  \mathcal D\Bar\psi_R^{\:\prime}\,\mathcal D\psi_L^{\prime} \;=\;
  \Bar J_R\,J_L\,\mathcal D\Bar\psi_R\,\mathcal D\psi_L\;,
\end{equation}
with
\begin{equation}
  \quad \Bar J_R \;\equiv\; \big(\det V^{\rm T}_+[\varphi/2]\big)^{-1}\;,
  \qquad J_L     \;\equiv\; \big(\det V_-[\varphi/2]\big)^{-1}\;.
\end{equation}
For the moment, the Jacobians in Eq.~\refeq{pidt} are only considered as
formal expressions.

The regularized determinant arising from a phase transformation \refeq{redef0}
with an infinitesimal parameter $\varphi(x)$ $=$
$\alpha(x)$ can be calculated using the plane-wave method of Ref. \cite{F80}.
The result is
\begin{equation}\label{detValpha}
  \det V_\pm[\alpha] \;=\; \Exp{i \, \mathcal{A}_{\pm}[\alpha,M]}\;,
\end{equation}
with
\begin{equation}
  \mathcal{A}_\pm[\alpha,M] \;\equiv\; \frac{M^2}{4\pi\;}
  \int_{T^{2} [i]}\rmd^2x\;\alpha(x)\,\,\e^{\pm
  F(x)/M^2}\;,
\end{equation}
where the regulator mass $M$ is to be taken to infinity at the end of the calculation.
Equation \refeq{detValpha} will be adopted as the proper
definition of the determinant of the matrices $V_\pm[\alpha]$ for infinitesimal
$\alpha(x)$. For later convenience, we establish two further identities:
\begin{equation}\label{i1}
  \det V_\pm[\alpha] \det V_\pm[\beta] =
  \det V_\pm[\alpha+\beta]\;,
\end{equation}
and
\begin{equation}\label{i2}
    \det V_\pm^{\rm T}[\alpha] = \det V_\pm^*[-\alpha] = \Big(\det V_\pm[-\alpha]\Big)^* =
    \det V_\pm[\alpha]\;,
\end{equation}
for infinitesimal $\alpha(x)$ and $\beta(x)$.

\subsection{Chiral determinant}

The method used in the previous subsection holds for infinitesimal phase
transformations. Here, we simply \emph{define} the
determinant of a finite phase transformation \refeq{redef0} to be
\begin{equation}\label{i3}
  \det V_\pm[\varphi/2] \equiv \lim_{N\rightarrow \infty}
  \Big( \det V_\pm[\varphi/(2N)]\Big)^N\;.
\end{equation}
For topologically trivial functions $\varphi(x)$ $=$ $\omega(x)$
as given in Eq. \refeq{varphi},
this definition is un\-prob\-lem\-at\-ic.
For topologically nontrivial functions $\varphi(x)$ $=$ $\chi(x)$
as given in Eq. \refeq{mn}, on
the other hand, the result turns out to break translation invariance; cf. Eq.
\refeq{zweibeinbcs}. The corresponding phase factor is, nevertheless, well-behaved
for the appropriate limit $M$ $\rightarrow$ $\infty$, as will become clear shortly.

 From Eqs. \refeq{detValpha} and \refeq{i3}, the combined regularized Jacobian
\refeq{pidt} for a left-moving fermion and its conjugate is found to be given by
\begin{equation}\label{theta}
    \Bar J_R \,J_L=  \big(\det V^{\rm T}_+[\varphi/2] \big)^{-1}\,
                   \big(\det V_-[\varphi/2]  \big)^{-1}
                = \Exp{-i W[\varphi,F,M]}\;,
\end{equation}
with
\begin{equation}\label{Theta0}
  W[\varphi,F,M] \;\equiv\;  \frac{M^2}{4\pi\;} \int_{T^{2} [i]}\!\rmd^2x\;
    \varphi(x)\,\cosh\!\big(F(x)/M^2\big)\;,
\end{equation}
and $F$ as defined in Eq. (\ref{Fdef}). For the topologically nontrivial part
$\chi(x)$ $=$ $(2\pi/L)$ $(m\,x^1+n\,x^2)$ of $\varphi(x)$,
the corresponding phase factor \refeq{theta}
approaches $1$ for $M^2$ $=$ $8\pi N/L^2$ with integer $N$ $\rightarrow$ $\infty$.
The remaining (trans\-la\-tion-in\-vari\-ant) phase factor depends only on the
topologically trivial part $\omega(x)$ of $\varphi(x)$.

We can now express the chiral determinant \refeq{De} for nontrivial zweibeins
($e^a_\mu \neq \id$) in terms of the
chiral determinant for trivial zweibeins ($e^a_\mu = \id$):
\begin{equation}\label{D0}
  D^{\{\theta_1,\theta_2\}}[A,e] =  \Exp{i W[\varphi,F,M]}\;
  D^{\{\theta_1',\theta_2'\}}[A,\id]\;,
\end{equation}
with the definitions
\begin{equation}\label{phit0}
  2\,\theta_1' \;\equiv\; (2\,\theta_1 + m) \text{ mod } 2\;,
  \qquad 2\,\theta_2' \;\equiv\; (2\,\theta_2 + n) \text{ mod } 2
\end{equation}
and the understanding that $M$ has to be taken to infinity in the way discussed in the
previous paragraph. In fact,
the regulator dependence drops out in the limit $M\,\rightarrow\,\infty$ for the
physically relevant ratio of chiral determinants:
\begin{equation}\label{ratio}
  \frac{D^{\{\theta_1,\theta_2\}}[A,e]}{D^{\{\theta_1,\theta_2\}}[B,e]}
  \;=\; \frac{D^{\{\theta_1',\theta_2'\}}[A,\id]}{D^{\{\theta_1',\theta_2'\}}[B,\id]}\;,
\end{equation}
with modified  spinor boundary conditions  given by the parameters
$\theta_1'$ and $\theta_2'$ of Eq.~\refeq{phit0}.
Here, $B$ is considered to be a fixed reference field, for example
$B_1(x)$ $=$  $B_2(x)$ $=$ $(2\pi/L)/\sqrt{2}\,$
[this particular choice is motivated by Eq. (\ref{chdt}) below].
Equation \refeq{ratio} is the main result of this section.

Two remarks on the result \refeq{ratio} are in order. The first remark is
that, in the end, only the topologically nontrivial part $\chi(x)$
of $\varphi(x)$ contributes to this ratio of  chiral determinants, i.e.
the dependence on the function $\omega(x)$ from Eq.~\refeq{varphi} drops out.
The second remark is that the torsion does not affect the translation invariance of
the normalized Euclidean effective action (defined as
minus the logarithm of the chiral determinant),
because the \rhs~of Eq.~\refeq{ratio} is translation-invariant by construction \cite{IN99}.

\subsection{Heuristic argument}

We now present a heuristic argument \cite{KN01}
for the change of the chiral determinant due to the
presence of torsion, restricting ourselves to the case of a constant gauge potential
\begin{equation}\label{constA}
  A_\mu=(2\pi/L)\,h_\mu
\end{equation}
and constant torsion tensor (\ref{torsion}) determined by
\begin{equation}\label{tildephi}
  \varphi(x) = \chi(x) = (2\pi/L)\,(m\,x^1+n\,x^2)\;.
\end{equation}
The path integral \refeq{De} to be calculated is then
\begin{equation}\label{pathint}
  D^{(\theta_1,\theta_2)}[h_\mu,m,n] \;\equiv\;
  \int \left[\mathcal D\Bar\psi_R\,\mathcal D\psi_L\right]_{(\theta_1,\theta_2)}\,
  \Exp{-\mathcal S[\Bar\psi_R,\psi_L,h_\mu,m,n]}\;,
\end{equation}
with the simplified action
\begin{equation}
  \mathcal S[\Bar\psi_R,\psi_L,h_\mu,m,n]
  =\int_{T^{2} [i]}\rmd^2 x\;\Bar\psi_R(x)\,
  \e^{i\chi(x)/2} \sigma^\mu
  (\partial_\mu+i\, 2\pi h_\mu/L) \e^{i\chi(x)/2}\psi_L(x)\;.
\end{equation}

The one-component spinors $\psi_L$ and $\Bar\psi_R$ with boundary conditions determined by
$\theta_1$ and $\theta_2$ can be expressed in a Fourier basis as follows:
\begin{equation}
  \begin{split}
    \psi_L &\;\equiv\; \sum_{p_1,p_2\in\Zset} b_{p_1p_2}\,\Exp{+
      \frac{2\pi i}{L}\,\Big(\big(p_1+\half+\theta_1\big)\,x^1+
                             \big(p_2+\half+\theta_2\big)\,x^2
                         \Big) }\;,\\[0.25\baselineskip]
    \Bar\psi_R &\;\equiv\; \sum_{q_1,q_2\in\Zset} \Bar a_{q_1q_2}\,\Exp{-
      \frac{2\pi i}{L}\,\Big(\big(q_1+\half+\theta_1\big)\,x^1+
                             \big(q_2+\half+\theta_2\big)\,x^2
                         \Big) }\;,\\
  \end{split}
\end{equation}
with Grassmann variables $b_{p_1p_2}$ and $\Bar a_{q_1q_2}$.
The measure of the path integral \refeq{pathint} can be written as in
Eq.~\refeq{measure}, but with $k$ and $l$ replaced by pairs of integers
$(q_1,q_2)$ and $(p_1,p_2)$.

Using this plane wave
decomposition, the path integral of Eq.~\refeq{pathint} is formally given by the
following infinite product:
\begin{equation}\label{chdt}
  \prod_{p_1,p_2\in\Zset} \Big(p_1+1/2+\theta_1+m/2+h_1
                               + i\big(p_2+1/2+\theta_2+n/2+h_2\big)\Big)\;,
\end{equation}
up to a constant overall factor. Although this expression needs
regularization, we can already infer that the introduction of torsion only changes the
spinor boundary conditions. Namely, $m/2$ appears together with $\theta_1$ and $n/2$
with $\theta_2$. Hence, the chiral determinant of a $U(1)$ gauge theory on
$T^{2} [i]$,
with torsion determined by $\varphi(x)=\chi(x)$ and with constant gauge potentials, is
proportional to
the chiral determinant of the theory without torsion and new spinor boundary
conditions $\theta_1'$ and $\theta_2'$ given by Eq.~\refeq{phit0}.  This explains the
result found in the previous subsection, at least for the particular gauge
potentials \refeq{constA} and torsion parameter function \refeq{tildephi}.

In Appendix~A, we discuss the zeta-function
regularization of a product similar to the one of
Eq.~(\ref{chdt}), which occurs for the vector-like $U(1)$ gauge theory. Again, the
spinor boundary conditions are found to be changed according to Eq.~\refeq{phit0}.

\section{CPT anomaly with torsion}

The calculation of Section~4.2  has shown how to relate the chiral determinant with
torsion to the standard chiral determinant without torsion.
The result \refeq{ratio} demonstrates that the
introduction of the topologically nontrivial zweibeins \refeq{zb}--\refeq{mn}
can effectively change the spinor boundary
conditions according to the constants $m$ and $n$ appearing in the torsion tensor
\refeq{torsion}.

It has been shown in Ref.~\cite{KN01} that the CPT anomaly
of  chiral $U(1)$ gauge theory on the torus without torsion
appears only for
\emph{doubly-periodic} spinor boundary conditions, at least for a particular class of
regularizations that respect modular invariance.
Under a CPT transformation of the gauge potential,
\begin{equation}
  A_\mu(x)\quad\longrightarrow
  \quad A_\mu^{\text{CPT}}(x)\;\equiv\;-A_\mu(-x)\;,
\end{equation}
the CPT anomaly on $T^{2} [i]$ manifests itself
as a sign change of the chiral determinant,
\begin{equation}\label{CPTanomaly}
  D^{\{1/2,1/2\}}[A^{\rm CPT},\id] = -D^{\{1/2,1/2\}}[A,\id]\;,
\end{equation}
for the case of trivial zweibeins ($e_\mu^a =\delta^a_\mu \equiv \id\,$).

In this paper, we only consider a \emph{single} charged chiral fermion. The
chiral $U(1)$ gauge anomaly needs, however, to be cancelled between different
species of chiral fermions. There is then the CPT anomaly (\ref{CPTanomaly}),
as long as the \emph{total} number $N_F$ of charged chiral fermions is \emph{odd}.
Note that even if there is no net CPT anomaly (that is, for $N_F$ even),
there may still be Lorentz noninvariance; see Ref.~\cite{KN01} for further details.

A consequence of our result \refeq{D0} is that the CPT anomaly can be moved to different
spinor boundary conditions by choosing appropriate zweibeins.
(The additional phase factor \refeq{theta} is CPT-even.)
For example, we can now have the CPT
anomaly for \emph{doubly-antiperiodic} spinor boundary conditions,
\begin{equation}\label{CPTAEx}
  D^{\{0,0\}}[A^{\rm CPT},\Bar e] = -D^{\{0,0\}}[A,\Bar e]\;,
\end{equation}
provided the zweibeins $\Bar e_\mu^{\,a}(x)$ have topologically nontrivial
torsion determined by \emph{odd} constants $m$ and $n$ in Eq.\ \refeq{mn}.

According to the heuristic argument of Section~4.3,
the chiral determinant is formally proportional to the infinite product~\refeq{chdt}.
It is then easy to understand  that there is a CPT anomaly if
both $2\,\theta_1+m$ and $2\,\theta_2+n$ are odd.
Start, for example, with purely
antiperiodic spinor boundary conditions. Now, the introduction of torsion with odd $m$ and
odd $n$ formally leads to the infinite product
\begin{equation}
 \prod_{p^{\,\prime}_1,p^{\,\prime}_2\in\Zset} (p^{\,\prime}_1+h_1+ip^{\,\prime}_2+ih_2)\;,
    \end{equation}
which equals the chiral determinant of a torsionless theory
with doubly-pe\-ri\-od\-ic spinor boundary
conditions.  Under a CPT transformation, $h_\mu\rightarrow-h_\mu$, the single factor with
$p^{\,\prime}_1=p^{\,\prime}_2=0$ is CPT-odd,
whereas the other factors combine into a CPT-even product (which
still needs to be regularized). Hence, for torsion determined by odd $m$ and odd
$n$, the CPT anomaly has been moved to the doubly-antiperiodic spin structure.
Analogous arguments apply to the other cases.

To summarize,
the CPT anomaly for chiral $U(1)$ gauge theory with an odd  number of charged chiral
fermions on the torus $T^{2} [i]$ occurs only if the following conditions hold:
\begin{equation}\label{conditions}
 (2\,\theta_1 + m)= 1 \text{ mod } 2\;, \qquad (2\,\theta_2 + n)= 1 \text{ mod } 2\;,
\end{equation}
at least for the regularizations used in Refs.~\cite{KN01,IN99}.
Here, $\theta_1$ and $\theta_2$ determine the fermion \bcs~\refeq{bcs}
and $m$ and $n$ are constants appearing in the topologically nontrivial zweibeins
\refeq{zb}--\refeq{mn}.

\section{Discussion}

For two-dimensional chiral $U(1)$ gauge theory, we have presented in this paper
a calculation of the chiral determinant on the torus
$T^{2} [i]$ with nontrivial zweibeins
corresponding to the presence of torsion on the spacetime manifold.

In Section~4.2 we have shown how to relate the chiral determinant with torsion to the chiral
determinant without torsion by the spinor redefinition \refeq{redef0}. The Jacobian
of this redefinition turns out to be a gauge-invariant and CPT-even phase factor
\refeq{theta}, which cancels in the ratio of the chiral determinants \refeq{ratio}.
The chiral determinant with torsion is then proportional to the chiral determinant without
torsion, but with spinor boundary conditions changed according to Eq.~\refeq{phit0}. This
result was confirmed  in Section~4.3
by a heuristic argument for a particular choice of gauge potentials and
torsion.
Hence, the CPT anomaly can effectively be moved from one spin structure to another
by choosing topologically nontrivial zweibeins.

The calculations of the present paper demonstrate that the two-dimensional
CPT anomaly is a genuine effect for chiral $U(1)$ gauge theory on the torus.
The CPT anomaly can be moved around between the
different spin structures by taking appropriate zweibeins; see Eq.~(\ref{conditions}).
But the anomaly cannot be removed completely from the general theory, which is
a sum over all spin structures \cite{AI79}.

Alternatively,
we can fix the spinor boundary conditions (for example,
antiperiodic boundary conditions \refeq{bcs} with $\theta_1$ $=$ $\theta_2$ $=$ $0$)
and consider different classes ($m,n \in \mathbb Z$) of zweibeins
(\ref{zb})--(\ref{mn}), with the corresponding torsion tensor (\ref{torsion}).
It is quite remarkable that topologically nontrivial spacetime torsion,
which is not visible in the metric and the curvature,
can affect the local physics of chiral $U(1)$ gauge theory
in the same way as different spinor boundary conditions would do for the
case of trivial zweibeins (i.e. vanishing torsion).

But, as we have shown in Section~2, there are more consequences of
torsion than just modified spinor boundary conditions.
There is, for example, the fact that
parallelograms do not close and that extremal and autoparallel curves need not coincide
if torsion is present. Moreover, these local manifestations of torsion can already occur
for topologically trivial zweibeins with $m=n=0$
in Eq.~(\ref{mn}), whereas the boundary-like effects require
topologically nontrivial zweibeins ($m\neq0$ or $n\neq0$).
Still, topologically nontrivial zweibeins may have a special status, as
discussed in Section~2.4.

In this paper, we have focused on two-dimensional chiral $U(1)$
gauge theory, because the chiral determinant is known exactly \cite{IN99}.
But our discussion of the effects of torsion can be readily
extended to higher-dimensional orientable manifolds.
Consider, for example, the flat spacetime manifold
$\mathbb R^{\,2}\times T^{2}$, with noncompact coordinates
$x^0,x^3$ $\in$ $\mathbb R$ and periodic coordinates $x^1,x^2$ $\in$ $[0,L]$.
The zweibeins \refeq{zb}--\refeq{mn} can then be embedded in the vierbeins $e^A_M(x)$ as
follows:
\begin{equation}\label{eAM}
 e^A_M(x) = \left\{
\begin{array}{ll}
e^a_\mu(x)\;, &\quad\mathrm{for}\;  A=a   \in    \{1,2\} \;,\;  M=\mu \in    \{1,2\} \;, \\
\delta^A_M\;, &\quad\mathrm{otherwise} \;,
\end{array} \right.
\end{equation}
with indices $A$ and $M$ running over 0, 1, 2, 3.
The $SO(2)$ angle $\varphi(x)$ which enters the nontrivial zweibein part of
Eq. \refeq{eAM} is taken to be purely topological, namely
$\varphi(x)$ $=$ $\chi(x^1,x^2)$ with $\chi$ as given by Eq. \refeq{mn}.

The metric  resulting  from the vierbeins \refeq{eAM} is flat, $g_{MN}(x)=\delta_{MN}$.
Note, however,  that the vierbeins \refeq{eAM}  with $\chi\neq 0$ do not
solve the vacuum field equations of the Einstein--Cartan
theory \cite{HHKN76}, in contrast with the situation in two dimensions as
mentioned in Footnote \ref{ftn1}. These vierbeins could play the role of
prior-geometric fields (that is, non-dynamical fields); see, for example,
the discussion in Ref. \cite{W93}. For the moment, let us just continue with the
particular vierbeins \refeq{eAM}, regardless of their origin.

In order to be specific, we also take the particular chiral gauge theory corresponding
to the  well-known $SO(10)$ grand-unified theory with three families of quarks and
leptons. The CPT anomaly now gives two Chern--Simons-like terms \cite{K00}
for the hypercharge $U(1)$ gauge field in the effective action, again provided
condition \refeq{conditions} holds.\footnote{It has been claimed in Ref.
\cite{DM97} that a cosmic torsion field
$S_\mu(x)$ $\equiv$ $\epsilon_{\mu\nu\rho\sigma}\,T^{\nu\rho\sigma}(x)$
could also generate a \CS-like term for the photon field via the quantum
effects of Dirac fermions coupled to both photon and torsion fields.
This radiatively induced \CS-like term must, however, vanish according to an
argument based on gauge invariance and analyticity \cite{CG99,P01} or, for
constant $S_\mu$ in particular, causality \cite{AK01plb}.
Note that the CPT anomaly necessarily involves chiral (Weyl) fermions,
not Dirac fermions \cite{K00,KN01}.} These Chern--Simons-like terms affect the
local physics, making the propagation of photons birefringent for example
\cite{CFJ90,AK01npb}.
This last phenomenon is all the more remarkable, since
at tree level torsion does not couple to the photons because of
gauge invariance  \cite{HHKN76}.

To summarize, a topological component of a (prior-geometric) torsion field
could modify the propagation of photons via the CPT anomaly.
Inversely, the propagation of photons could perhaps inform us about the
structure of spacetime.

\appendix
\section{Dirac determinant for two-dimensional $U(1)$ gauge theory with torsion}

In this appendix, we evaluate the regularized fermionic determinant of a
two-di\-men\-sion\-al
$U(1)$ gauge theory with a single Dirac fermion, i.e. the vector-like $U(1)$ gauge
theory. The spacetime manifold considered is the torus $T^{2}[i]$ shown in
Fig.~1. In order to simplify the calculation, we take, as in Section~4.3,
constant gauge potentials $A_\mu(x)=(2\pi/L)\,h_\mu$, with $h_\mu$ $\in$
$\mathbb R$, and constant torsion tensor components \refeq{torsion} determined by
$\varphi(x) =\chi(x) = (2\pi/L)\,(m\,x^1+n\,x^2)$, with $m,n$ $\in$ $\Zset$.

For a single Dirac fermion, the fermionic determinant
(exponent of minus the Euclidean effective action) is the infinite product of the
following eigenvalues:
\begin{equation}\label{lambdap1p2}
  \lambda_{p_1p_2} \equiv
  (2\pi/L)^2\, \big((p_1+a_1)^2+(p_2+a_2)^2\big)\;,
\end{equation}
with quantum numbers $p_1,p_2\in\Zset$ and (noninteger) parameters
\begin{equation}
  a_1 \;\equiv\; 1/2+\theta_1+m/2+h_1\;,\qquad
  a_2 \;\equiv\; 1/2+\theta_2+n/2+h_2\;.
\end{equation}
Compare with the product (\ref{chdt}) for a single chiral fermion.

This product of eigenvalues can be regularized using
zeta-function techniques (see Refs.~\cite{H77,BVW91} and references therein).
For $\vec a \equiv (a_1,a_2)$, we define the regularized Dirac determinant as follows:
\begin{equation}\label{Dzeta}
 D_{\,\mathrm{Dirac}}^{\,\{\theta_1,\theta_2\}}[h_\mu,m,n]
 \;\equiv\; \Exp{-\zeta^{\,'}_E(s,\vec a)\big|_{s=0}}\;,
\end{equation}
with the generalized Epstein zeta function
\begin{equation}\label{gEZ}
  \zeta_E(s,\vec a) \;\equiv\; \sum_{p_1,p_2\in\Zset}
  \big((p_i+a_i)g^{ij}(p_j+a_j)\big)^{-s}\;,
\end{equation}
for $g^{ij} \equiv (2\pi/L)^2\,\delta^{ij}$ and $\mathrm{Re}\,(s)$ $>$ $1$.
The prime in Eq. (\ref{Dzeta}) denotes differentiation with respect to the
variable $s$ (which is set to 0 afterwards).

Our evaluation of the sum \refeq{gEZ} essentially repeats the calculation of
Ref. \cite{BVW91}, to which the reader is referred for further details.
In the rest of this appendix, $g_{ij}$ will stand for the
inverse of the matrix $g^{ij}$ and we will set $p^i\equiv p_i$.

By writing the generalized Epstein zeta function \refeq{gEZ} as a Mellin transform,
\begin{equation}\label{zE}
  \zeta_E(s,\vec a) = \frac{1}{\Gamma(s)}\, \sum_{p_1,p_2\in\Zset}
  \;\int_0^\infty\rmd t\; t^{s-1}\Exp{-t\,\lambda_{p_1p_2}}\;,
\end{equation}
we can apply the generalized Poisson resummation formula,
\begin{eqnarray}
  &&\sum_{p_1,p_2\in\Zset}\Exp{-\pi (p_i+a_i)g^{ij}(p_j+a_j)} =\nonumber\\[0.25\baselineskip]
  &&\sqrt{\det (g_{ij})}\,
  \sum_{p_1,p_2\in\Zset}\Exp{-\pi\, p^ig_{ij}\,p^j + 2\pi i\,p^ja_j}\;,
\end{eqnarray}
to the integrand of Eq.~\refeq{zE}. The result is given by
\begin{equation}\label{nozero}
  \zeta_E(s,\vec a) = \frac{\Gamma(1-s)}{\Gamma(s)}\, \pi^{s-1}\, \sqrt{\det(g_{ij})}\,
  \sum_{p_1,p_2\in\Zset}\!\!\!\!\!\!\ ^{'} (p^i g_{ij}\,p^j )^{s-1}
  \Exp{2\pi i\,p^j a_j}\;,
\end{equation}
with the prime on the sum indicating that the
modes $p_i=0$ are excluded, since they do not contribute for the region
$\mathrm{Re}\,(s)>1$ where the original
sum \refeq{gEZ} is convergent. Analytic continuation to $s=0$ then yields
\begin{equation}
  \begin{split}
    \zeta_E(0,\vec a) &= 0\;,\\[0.25\baselineskip]
    \zeta_E^{\,'}(0,\vec a) &= \pi^{-1}\, \sqrt{\det(g_{ij})}\,
    \sum_{p_1,p_2\in\Zset}\!\!\!\!\!\!\ ^{'} (p^i g_{ij}\, p^j )^{-1}
    \Exp{2\pi i\,p^ja_j}\;.
  \end{split}
\end{equation}

It is a remarkable fact \cite{BVW91} that one can express this
$\zeta_E^{\,'}(0,\vec a)$ in
terms of the Riemann theta function and Dedekind eta function
\begin{equation}\label{detH}
    \zeta_E^{\,'}(0,\vec a) = -\log\left|
    \frac{1}{\eta(\tau)}\;\THETA{1/2-a_1}{1/2+a_2}(0,\tau) \right|^2 \;,
\end{equation}
with modulus $\tau=i$ for the particular matrix $g^{ij}$ $\propto$ $\delta^{ij}$ of
Eq.~(\ref{gEZ}).
Here, the Riemann theta function with characteristics $a$ and $b$ is defined as in
Ref.~\cite{M83},
\begin{equation}\label{tab}
  \THETA{a}{b}(z,\tau)\;\equiv\;\sum_{n\in\Zset}\Exp{i\pi\tau(n+a)^2+2\pi i (n+a)(z+b)}\;,
\end{equation}
and the Dedekind eta function is given by
\begin{equation}
  \eta(\tau)\;\equiv\; \e^{i\pi\tau/12}\;\prod_{m=1}^\infty(1-\e^{2\pi i\tau m})\;.
\end{equation}
Note that the regularization method used has eliminated the $L$-dependence
present in Eq.~\refeq{lambdap1p2} and produced the result \refeq{detH}
which does not depend on $L$; cf. Refs. \cite{H77,BVW91}.

The theta functions \refeq{tab} obey the following identity:
\begin{equation}\label{idTH}
  \THETA{a+N}{b+M}(z,\tau) = \e^{2\pi i aM}\;\THETA{a}{b}(z,\tau)\;,
\end{equation}
for arbitrary integers $N$ and $M$. In addition, there are some further properties for
the special case of $z=0$ and $\tau=i$, which allow us to write Eqs.~\refeq{Dzeta}
and \refeq{detH} as
\begin{equation}\label{logdetHmn}
   \log D_{\,\mathrm{Dirac}}^{\,\{\theta_1,\theta_2\}}[h_\mu,m,n] =
  -\zeta_E^{\,'}(0,\vec a) =\log\left| \frac{1}{\eta(i)}\;
  \THETA{\theta_1'+h_1}{\theta_2'+h_2}(0,i) \right|^2 \;,
\end{equation}
with
\begin{equation}\label{Aphip}
  2\,\theta_1'  \;\equiv\; (2\,\theta_1 + m) \mathrm{\ mod\ } 2\;,\qquad
  2\,\theta_2'  \;\equiv\; (2\,\theta_2 + n) \mathrm{\ mod\ } 2\;.
\end{equation}
This shows that the effect of torsion (parameters $m$ and $n$)
for the regularized Dirac determinant can be
entirely absorbed by a change of spinor boundary conditions, as given by Eq.
(\ref{Aphip}). Note also that the identity \refeq{idTH} implies
the gauge invariance of \refeq{logdetHmn} under
$h_\mu$ $\rightarrow$ $h_\mu + n_\mu$, for $n_\mu$ $\in$ $\mathbb Z$.


\begin{thebibliography}{999}
\bibitem{K00}    F.R. Klinkhamer,
                 {\it Nucl. Phys.} {\bf B 578} (2000) 277.

\bibitem{P55}    W. Pauli, in: W. Pauli, L. Rosenfeld and V. Weisskopf, eds.,
                 {\it Niels Bohr and the Development of Physics}
                 (Pergamon, London, 1955) p. 30.

\bibitem{L57}    G. L\"{u}ders,  {\it Ann. Phys. (N. Y.)} {\bf 2} (1957) 1.

\bibitem{K98}    F.R. Klinkhamer,
                 {\it Nucl. Phys.} {\bf B 535} (1998) 233.

\bibitem{KN01}   F.R. Klinkhamer and J. Nishimura,
                 {\it Phys. Rev.} {\bf D 63} (2001) 097701.

\bibitem{IN99}   T. Izubuchi and J. Nishimura,
                 {\it J. High Energy Phys.} {\bf 10} (1999) 002.

\bibitem{HHKN76} F.W. Hehl, P. von der Heyde, G.D. Kerlick and J.M. Nester,
                 {\it Rev. Mod. Phys.} {\bf 48} (1976) 393.

\bibitem{EGH80}  T. Eguchi, P.B. Gilkey and A.J. Hanson,
                 {\it Phys. Rep.} {\bf 66} (1980) 213.

\bibitem{AI79}   S.J. Avis and C.J. Isham,
                 {\it Nucl. Phys.} {\bf B 156} (1979) 441.

\bibitem{BD79}   R. Banach and J.S. Dowker,
                 {\it J. Phys.} {\bf A 12} (1979) 2545.

\bibitem{E55}    A. Einstein, {\it The Meaning of Relativity}, 5-th ed.
                 (Princeton University Press, Princeton, 1955) pp. 70, 145.

\bibitem{DB79}   D.H. Douglass and V.B. Braginsky, in: S.W. Hawking and W. Israel,
                 eds., {\it General Relativity: An Einstein Centenary Survey}
                 (Cambridge University Press, Cambridge, 1979) p. 90.

\bibitem{F80}    K. Fujikawa,
                 {\it Phys. Rev.} {\bf D 21} (1980) 2848; {\bf D 29} (1984) 285.

\bibitem{W93}    C. Will, {\it Theory and Experiment in Gravitational Physics},
                 2-nd ed. (Cambridge University Press, Cambridge, 1993) p. 17.

\bibitem{DM97}   A. Dobado and A.L. Maroto,
                 {\it Mod. Phys. Lett.} {\bf A 12} (1997) 3003.

\bibitem{CG99}   S. Coleman and  S.L. Glashow, {\it Phys. Rev.} {\bf D 59} (1999) 116008.

\bibitem{P01}    M. P\'{e}rez-Victoria, {\it J. High Energy Phys.} {\bf 04} (2001) 032.

\bibitem{AK01plb}  C. Adam and F.R. Klinkhamer,
                 {\it  Phys. Lett.}  {\bf B 513} (2001) 245.

\bibitem{CFJ90}  S.M. Carroll, G.B. Field and R. Jackiw,
                 {\it Phys. Rev.} {\bf D 41} (1990) 1231.

\bibitem{AK01npb}   C. Adam and F.R. Klinkhamer,
                 {\it Nucl. Phys.} {\bf B 607} (2001) 247.

\bibitem{H77}    S.W. Hawking,
                 {\it Comm. Math. Phys.} {\bf 55} (1977) 133.

\bibitem{BVW91}  S.K. Blau, M. Visser and A. Wipf,
                 {\it Int. J. Mod. Phys.} {\bf A 6} (1991) 5409.

\bibitem{M83}    D. Mumford,
                 {\it Tata Lectures on Theta} (Birkh\"auser, Boston, 1983).

\end{thebibliography}
\end{document}